\documentclass[apj,numberedappendix]{emulateapj}
\usepackage{apjfonts} 

\makeindex
\citeindextrue
 
\def\deg{\ifmmode^\circ\else$^\circ$\fi}
\def\min{\ifmmode^\prime\else$^\prime$\fi}

\shorttitle{Relativistic beaming in radio jets}
\shortauthors{Cohen et al.}
\received{October 6, 2006}
\revised{November 15, 2006}
\accepted{November 16, 2006}
\submitted{}
\journalinfo{Astrophysical~journal}

\defcitealias{kik98}{Paper~I}
\defcitealias{Z02}{Paper~II}
\defcitealias{Kellermann_etal04}{Paper~III}
\defcitealias{Ketal05}{Paper~IV}
\defcitealias{LM97}{LM97}
\defcitealias{VC94}{VC94}
\defcitealias{J05}{J05}

\begin{document}

\title{RELATIVISTIC BEAMING AND THE INTRINSIC PROPERTIES 
OF EXTRAGALACTIC RADIO JETS 
}


\author{M. H. Cohen\altaffilmark{1},
        M. L. Lister\altaffilmark{2},
        D. C. Homan\altaffilmark{3},
        M. Kadler\altaffilmark{4,5}
        K. I. Kellermann\altaffilmark{6},
        Y. Y. Kovalev\altaffilmark{5,7,8},
        and
        R. C. Vermeulen\altaffilmark{9}
        }

\altaffiltext{1}{Department of Astronomy, Mail Stop 105-24,
                 California Institute of Technology,
                 Pasadena, CA 91125, U.S.A.;
                 \mbox{mhc@astro.caltech.edu}
                 }
\altaffiltext{2}{Department of Physics, Purdue University,
                 525 Northwestern Avenue, West Lafayette, IN
                 47907, U.S.A.;
                 \mbox{mlister@physics.purdue.edu}
                 }
\altaffiltext{3}{Department of Physics and Astronomy, Denison
University,
                 Granville, OH 43023, U.S.A.;
                 \mbox{homand@denison.edu}
                 }
\altaffiltext{4}{Astrophysics Science Division, NASA Goddard
                 Space Flight Center, Greenbelt Road, 
                 Greenbelt, MD 20771, USA;
                 \mbox{mkadler@milkyway.gsfc.nasa.gov}
                 }
\altaffiltext{5}{Max-Planck-Institut f\"ur Radioastronomie, 
                 Auf dem H\"ugel 69, 53121 Bonn, Germany;
                 \mbox{ykovalev@mpifr-bonn.mpg.de}
                 }
\altaffiltext{6}{National Radio Astronomy Observatory,
                 520 Edgemont Road, Charlottesville,
                 VA~22903--2475, U.S.A.;
                 \mbox{kkellerm@nrao.edu}
                 }
\altaffiltext{7}{Astro Space Center of Lebedev Physical
                 Institute,
                 Profsoyuznaya 84/32, 117997 Moscow, Russia;
                 }
\altaffiltext{8}{Jansky Fellow,
                 National Radio Astronomy Observatory,
                 P.O.~Box 2, Green Bank, WV 24944, U.S.A.;
                 }   
\altaffiltext{9}{ASTRON,
                 Netherlands Foundation for Research in
                 Astronomy,
                 P.O.~Box 2, NL-7990 AA Dwingeloo, The
                 Netherlands;
                 \mbox{rvermeulen@astron.nl}
                 }

\begin{abstract}

Relations between the observed quantities for a beamed radio jet, apparent
transverse speed and apparent luminosity ($\beta_\mathrm{app}$,$L$),
and the intrinsic quantities, Lorentz factor and intrinsic luminosity
($\gamma$,$L_o$), are investigated.  The inversion from measured to
intrinsic values is not unique, but approximate limits to $\gamma$
and $L_o$ can be found using probability arguments.  Roughly half
the sources in a flux density--limited, beamed sample have a
value of $\gamma$ close to the measured $\beta_\mathrm{app}$.
The methods are applied to observations of 119 AGN jets made with
the VLBA at 15~GHz during 1994--2002. The results strongly support
the common relativistic beam model for an extragalactic radio jet.
The ($\beta_\mathrm{app}$,$L$) data are closely bounded by a theoretical
envelope, an $aspect$ curve for $\gamma=32$, $L_o= 10^{25}$\,W\,Hz$^{-1}$.
This gives limits to the maximum values of $\gamma$ and $L_o$ in the
sample: $\gamma_\mathrm{max}\approx 32$, and $L_{o,\mathrm{max}}\sim
10^{26}$\,W\,Hz$^{-1}$.  No sources with both high $\beta_\mathrm{app}$
and low $L$ are observed.  This is not the result of selection effects
due to the observing limits, which are flux density $S>0.5$~Jy, and
angular velocity $\mu<4$ mas yr$^{-1}$.  Many of the fastest quasars
have a pattern Lorentz factor $\gamma_p$ close to that of the beam,
$\gamma_b$, but some of the slow quasars must have $\gamma_p\ll\gamma_b$.
Three of the 10 galaxies in the sample have a superluminal feature,
with speeds up to $\beta_\mathrm{app}\approx 6$. The others are at most
mildly relativistic.  The galaxies are not off--axis versions of the
powerful quasars, but \objectname{Cygnus~A} might be an exception.

\end{abstract}


\section{Introduction
}
\label{s:intro}

In recent years VLBI observations have provided many accurate values of
the apparent luminosity, $L$, of compact radio jets, and the apparent
transverse speed, $\beta_\mathrm{app}$, of features (components)
moving along the jets.  These quantities are of considerable interest,
but the intrinsic physical parameters, the Lorentz factor, $\gamma$,
and the intrinsic luminosity, $L_o$, are more fundamental. In this
paper we first consider the ``inversion problem;'' i.e., the
estimation of intrinsic quantities from observed quantities.
We then apply the results to data from a large multi--epoch survey
we have carried out with the VLBA at 15~GHz.

The inversion problem is discussed in \S\ref{s:beams}--\S\ref{s:inversion}
with an idealized relativistic beam, one that has the same vector
velocity everywhere, and contains a component moving with the beam
velocity. The jet emission is Doppler boosted, and Monte--Carlo simulations
are used to estimate the probabilities associated with selecting
a source: that of selecting ($\beta_\mathrm{app}$,$L$)
from a given ($\gamma$,$L_o$); and the converse, the probability
of ($\gamma$,$L_o$) being the intrinsic parameters for an observed
($\beta_\mathrm{app}$,$L$).

In \S\ref{s:inversion} we introduce the
concept of an ``aspect'' curve, defined as the track of a source on the
($\beta_\mathrm{app}$,$L$) plane (the observation plane) as $\theta$
(angle to the line-of-sight, LOS) is varied; and an ``origin'' curve,
defined as the set of values on the ($\gamma$,$L_o$) plane (the intrinsic
plane) from which the observed source can be expressed.  These provide a
ready way to understand the inversion problem, and illustrate the lack of
a unique inversion for a particular source.  Probabilistic limits provide
constraints on the intrinsic parameters for an individual source; but
when the entire sample of sources is considered, more general comments
can be made, as in \S\ref{s:distros}.

The observational data are discussed in \S\ref{s:data}.  They are from
a 2--cm VLBA survey, and have been published in a series of papers:
\citet[][hereafter \citetalias{kik98}]{kik98},
\citet[][\citetalias{Z02}]{Z02},
\citet[][\citetalias{Kellermann_etal04}]{Kellermann_etal04},
\citet[][\citetalias{Ketal05}]{Ketal05}, and E.~Ros et al.~2007, in
preparation. This is a continuing survey and the speeds are regularly
updated using new data; in this paper we include results up to 2006
September 15. The analysis also includes some results from the MOJAVE
program, which is an extension of the 2--cm survey using a statistically
complete sample \citep{LH05}.  Prior to \citetalias{Kellermann_etal04},
the largest compilation of internal motions was in \citet[][hereafter
\citetalias{VC94}]{VC94}, who tabulated the internal proper motion,
$\mu$, for 66 AGN, and $\beta_\mathrm{app}$ for all but the two without a
redshift. The data came from many observers, using various wavelengths
and different VLBI arrays, and consequently were inhomogeneous.
The 2--cm data used here were obtained with the VLBA over the period
1994--2002, and comprise ``Excellent'' or ``Good'' apparent speeds
\citepalias{Kellermann_etal04} for components in 119 sources. This is
a substantial improvement over earlier data sets, and allows us to make
statistical studies which previously have not been possible.  Other recent
surveys are reported by \citet[][hereafter \citetalias{J05}]{J05} with
data on 15~AGN at 43~GHz, \cite{Hom01} with data on 12~AGN at 15
and 22~GHz, and \cite{P06} with data on 77~AGN at 8~GHz.

In some sources it is clear that the beam and pattern speeds are
different, and to discuss this we differentiate between the Lorentz
factor of the beam, $\gamma_\mathrm{b}$, and that for the pattern,
$\gamma_\mathrm{p}$. In most of this paper, however, we assume
$\gamma_\mathrm{b} \approx \gamma_\mathrm{p}$ and drop the subscripts. In
\S\ref{s:Lorentz} and \S\ref{s:distros} peak values for the distributions
of $\gamma$ and $L_o$ in the sample are discussed, and in \S\ref{s:QB}
the low--velocity quasars and BL Lacs are discussed. It is likely that
some of these have components whose pattern speed is significantly less
than the beam speed. The radio galaxies in our sample are discussed in
\S\ref{s:galaxies}, and we show that most of them are not high--angle
versions of the powerful quasars.  Cygnus A may be an exception, and we
speculate that it contains a fast central jet (a spine), with a slow
outer sheath.

In this paper we use a cosmology with $H_0 = 70$~km\,s$^{-1}$\,Mpc$^{-1}$,
$\Omega_\mathrm{m}=0.3$, and $\Omega_\Lambda=0.7$.

\section{Relativistic Beams
}
\label{s:beams}

In this Section the standard relations for an ideal relativistic beam
\citep[e.g.,][]{BK79} are reviewed. The beam is characterized by its
Lorentz factor, $\gamma$, intrinsic luminosity, $L_o$, and angle $\theta$
to the line of sight. From these the Doppler
factor, $\delta$, the apparent transverse speed, $\beta_\mathrm{app}$,
and the apparent luminosity, $L$, can be calculated:

\begin{equation}
   \delta = \gamma^{-1}(1-\beta\cos\theta)^{-1}\,,
\label{eq:delta}
\end{equation}
\begin{equation}
   \beta_\mathrm{app} = \frac{\beta\sin\theta}{1-\beta\cos\theta}\,,
\label{eq:beta_app}
\end{equation}
\begin{equation}
   L = L_o \delta^n\,,
\label{eq:lum}
\end{equation}
where $\beta=(1-\gamma^{-2})^{1/2}$ is the speed of the beam
in the AGN frame (units of $c$) and $L_o$ is the luminosity that would
be measured by an observer in the frame of the radiating material. 
The exponent $n$ in equation~\ref{eq:lum} combines effects due to the
K correction and those due to Doppler boosting: $n = \alpha + p$, where
$\alpha$ is the spectral index ($S\sim\nu^\alpha$) and $p$ is the Doppler
boost exponent, discussed in \S\ref{s:assum}.

\begin{figure}[thb]
\centering
\resizebox{1.0\hsize}{!}
{
\includegraphics[angle=0,trim=1.5cm 0.5cm 0.8cm 0cm]{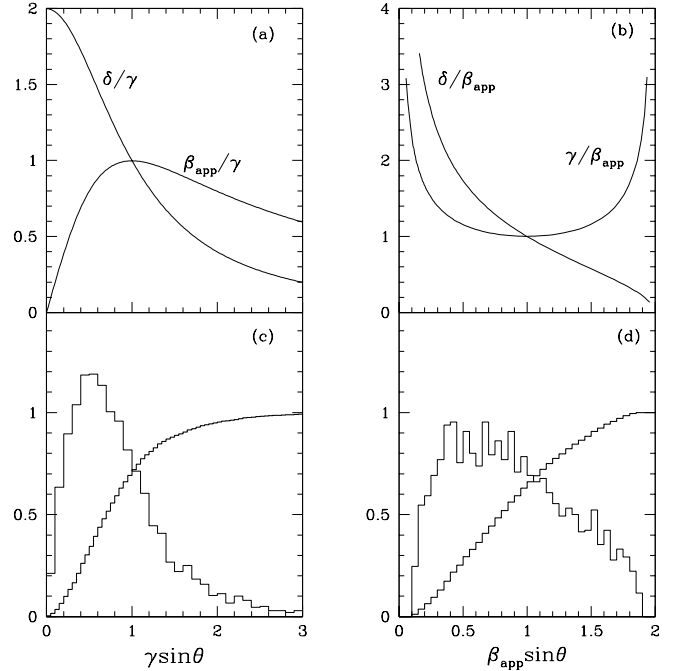} 
}
\caption{\label{f:beam_par}
$Top$: Parameters for a relativistic beam having Lorentz factor
$\gamma$ and angle to the LOS $\theta$.  (a) Curves are plotted for
$\gamma=15$ but change slowly with $\gamma$, provided $\gamma^2\gg 1$.
(b) Curves are plotted for $\beta_\mathrm{app}= 15$ and change slowly
with $\beta_\mathrm{app}$, provided $\beta_\mathrm{app}^2\gg 1$.
In (a) the quantities are normalized by the constant Lorentz factor;
in (b), by the constant apparent speed.  $Bottom$: Results from a
Monte--Carlo simulation of a flux density--limited survey selected
from the parent population described in Appendix~\ref{appendix}. (c)
Probability density $p(\theta|\gamma_\mathrm{f})$ and cumulative
probability $P(\theta|\gamma_\mathrm{f})$ (heavy line) for
$\gamma\approx 15$. Roughly 75\% of the selected sources will have
$\gamma\sin\theta<1$; i.e., $\theta<\theta_\mathrm{c}$. Values of
$\gamma\sin\theta < 0.15$ and $>2.0$
are unlikely; the cumulative probabilities are approximately 0.04 and
0.96, respectively.  (d) Probability density and cumulative probability
for selecting a source at angle $\theta$, for $\beta_\mathrm{app}\approx
15$. As $\beta_\mathrm{app}$ decreases the probability curve becomes
more peaked, and the peak moves to the left.
}
\end{figure}

From equations (\ref{eq:delta}) and (\ref{eq:beta_app}), any two of the four parameters
$\beta_\mathrm{app}$, $\gamma$, $\delta$, and $\theta$ can be used
to find the others; a convenient relation is $\beta_\mathrm{app}
= \beta\gamma\delta\sin\theta$. Figure~\ref{f:beam_par}a
shows $\delta$ and $\beta_\mathrm{app}$ as functions of
$\sin\theta$, all normalized by $\gamma$; the curves are
valid for $\gamma^2\gg 1$. When $\sin\theta=\gamma^{-1}$,
$\delta=\gamma$ and $\beta_\mathrm{app}=\beta_\mathrm{app,max}=
\beta\gamma$. The ``critical'' angle $\theta_\mathrm{c}$ is defined
by $\sin\theta_\mathrm{c}=\gamma^{-1}$, and the approximation
$\theta/\theta_\mathrm{c} \approx \gamma \sin\theta$ will be used; this
is accurate for $\gamma^2\gg 1$ and $\theta^2\ll 1$, and is correct to
20\% for $\theta<60\deg$ and $\beta>0.5$.

It is also useful to regard $\beta_\mathrm{app}$ and $\theta$ as the
independent quantities. Figure~\ref{f:beam_par}b shows $\delta$ and
$\gamma$ as functions of $\sin\theta$, all normalized by
$\beta_\mathrm{app}$. The curves are calculated for
$\beta_\mathrm{app}=15$ and change slowly with $\beta_\mathrm{app}$,
for $\beta_\mathrm{app}^2\gg 1$.

\subsection {Assumptions}
\label{s:assum}

In our analysis we assume that a source contains an ideal relativistic
beam; one that is straight and narrow, and in which the pattern speed is
the same as the beam speed: $\gamma_\mathrm{p}\approx\gamma_\mathrm{b}$.
In particular, the Doppler factor for the core must derive from
the same values of $\gamma$ and $\theta$ as apply to the value of
$\beta_\mathrm{app}$ for the moving component, several pc or more away.
Many sources, however, are seen to have more than one moving component,
and they may have different values of $\beta_\mathrm{app}$. In these
cases we have selected the fastest speed, on the grounds that of all the
components, it is the one most likely to be moving at near the beam speed.
It probably is due to a shock associated with an outburst in flux density,
while some of the slower components might be trailing shocks \citep{AGU01}.
The main shock itself must be moving faster than the beam, but the
synchrotron source, which is a density concentration behind the shock,
can have a net speed slower than the shock, as shown by numerical
simulations \citep{AGU01}.

In some sources the only component we see is stationary at a bend in the
jet. We believe that in these cases we see a standing shock, or perhaps
enhanced radiation from a section of the jet which is tangent to the
LOS. These components will have $\gamma_\mathrm{p}\ll\gamma_\mathrm{b}$
and might be part of the population of slow quasars discussed in
\S\ref{s:QB}.

It is  clear that some jets are not straight, and that $\theta$ is
not the same in the core and in the moving components. See, e.g.,
\objectname{3C~279} \citep{Hom03} where the velocity vector changed
during the course of observations, and \objectname{0735+178} and
\objectname{2251+158}, where the image shows a jet with sharp bends
\citepalias{kik98}. However, in cases where at least moderate
superluminal motion is found, the motion must be close to the LOS,
and any changes in angle will be strongly amplified by projection.
An observed right angle bend could correspond to an intrinsic
bend of only a few degrees.

The Doppler boost exponent $p$ depends on geometry and optical depth
and is discussed by \cite{LB85}. For a smooth jet, $p=2$, and this value
is appropriate for the core region, where, we assume, the relativistic
beam streams through a stationary $\tau=1$ region.  In some cases the
moving component can be modelled as an isolated optically thin source,
which would have $p=3$, but for most of the sources the flux density
is dominated by radiation from the core, and we use $p=2$ here. See
Figure~5 in \citetalias{Ketal05}.  The other term in
the exponent $n$ is the spectral index $\alpha$.  Nearly all the sources
have a ``flat" spectrum, $|\alpha|<0.5$, and they also have variable
flux density and a variable spectrum.  Because of the time dependence
it is not possible to generate a useful index for each source, and we
take $\alpha=0$ as a rough global average, giving $n=2$.  This is further
justified in \S\ref{s:boost_exp},  where it is shown that $n=3$ does not
fit the data. However, this choice of $\alpha=0$ clearly leads to error
for those sources with a high Doppler factor, say $\delta=30$, because
the K-correction must cover a frequency range of a factor of 30. This
introduces uncertainty into the estimates of intrinsic luminosity.

We shall use equation~(\ref{eq:lum}) as if $L_o$ is independent of
$\theta$, but this is not necessarily so. The opacity in the surrounding
material may change with $\theta$, and the luminosity of any optically
thick component may change with angle. Other changes in $L_o$ might be
caused by a change in location of the emission region, as $\theta$,
and therefore the Doppler factor and the emission frequency, change
\citep{Lob98}.

\section {Probability
}
\label{s:prob}

The probability of selecting a source with a particular value of
$\theta$, $\gamma$, $\beta_\mathrm{app}$, or $\delta$ from a flux
density--limited sample of relativistically--boosted sources is central
to our discussion.  Because $S\propto\delta^2$ (\S\ref{s:boost_exp})
and $\delta$ decreases with increasing $\theta$, the sources found
will preferentially be at small angles, even though there is not
much solid angle there. \citetalias{VC94} calculated the probability
$p(\theta|\gamma_f)$ (the subscript $f$ means {\em fixed}) in
a Euclidean universe, and \citet[][hereafter \citetalias{LM97}]{LM97}
extended this with Monte--Carlo calculations, to include evolution.
However, the observations directly give $\beta_\mathrm{app}$, not
$\gamma$, and $p(\theta|\beta_\mathrm{app,f})$ is generally not an
analytic function. To deal with this, M.~Lister et al. (in preparation)
are using Monte--Carlo methods to study the probability functions. We
use one of their simulations here, as an illustrative example.

In the Monte--Carlo calculation, a simulated parent population is
created (see Appendix~\ref{appendix}), from which one hundred
thousand sources with $S>1.5$\,Jy are drawn.  We select a
slice of this sample with $14.5\le\gamma\le 15.5$ and form the
histograms in Figure~\ref{f:beam_par}c, showing the probability
density $p(\theta|\gamma_\mathrm{f})$ and the cumulative probability
$P(\theta|\gamma_\mathrm{f})$ for those sources with $\gamma\approx 15$.
The histograms vary slowly with $\gamma$, provided $\gamma^2 \gg 1$.  They
are similar to the equivalent diagrams calculated by \citetalias{VC94}
(Figure~7) and by \citetalias{LM97} (Figure~5).  Figure~\ref{f:beam_par}c
may be directly compared with Figure~\ref{f:beam_par}a, which is a purely
geometric result from equations \ref{eq:delta} and \ref{eq:beta_app}. The
peak of the probability is at $\sin\theta\approx 0.6/\gamma$, where
$\beta_\mathrm{app} \approx 0.9\gamma$ and $\delta \approx 1.5\gamma$.
The 50\% point of $P(\theta | \gamma_\mathrm{f})$ is at $\gamma
\sin\theta\approx 0.7$, giving a median value $\theta_\mathrm{med}\approx
0.7\gamma^{-1}\approx 0.7\theta_\mathrm{c}$.

\begin{figure}[t]
\centering
\resizebox{0.9\hsize}{!}
{
\includegraphics[angle=0,trim=0.5cm 7cm 10.8cm 0.8cm]{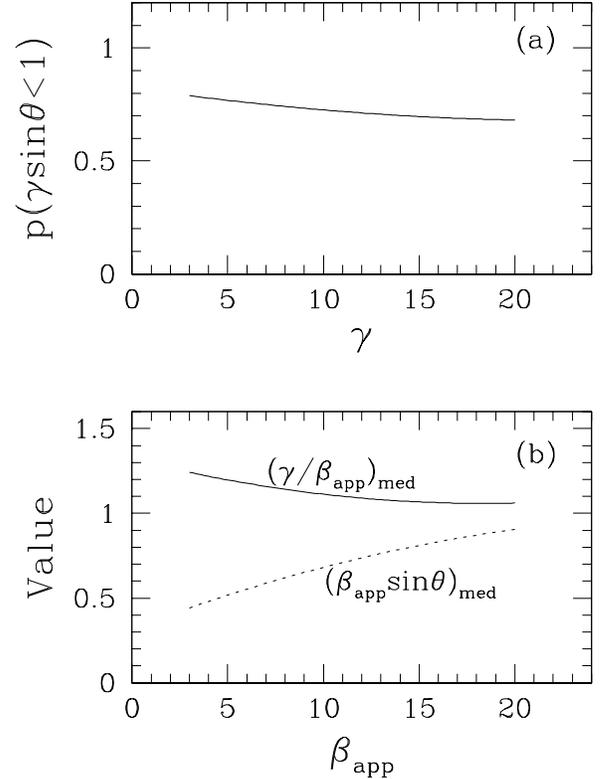}
}
\caption{\label{f:prob}
(a) Probability that $\gamma\sin\theta<1$ 
as a function of $\gamma$.  (b) (solid curve) median value of
$\gamma/\beta_\mathrm{app}$ and (dashed curve) median value of
$\beta_\mathrm{app}\sin\theta$, as functions of $\beta_\mathrm{app}$.
}
\end{figure}

An interesting measure of the cumulative probability is
$P(\theta=\theta_\mathrm{c})$, the fraction of the sample lying inside
the critical angle. The slow variation of this fraction with $\gamma$ is
seen in Figure~\ref{f:prob}a; a rough value is 0.75; i.e., most
beamed sources will be inside their ``$1/\gamma$ cones."  In this paper
we will take $0.04<P<0.96$ as a practical range for the probability.
This corresponds, approximately, to $0.15<\theta/\theta_\mathrm{c}<2$
for $\gamma=15$, and the angular range for this probability range varies
slowly with $\gamma$. Figure~\ref{f:appendix} (Appendix~\ref{appendix})
shows the $(\gamma,\theta)$ distribution for 14,000 sources from the
simulation, along with the 4\% and 96\% limits.

We have now described $p(\theta|\gamma_\mathrm{f})$, the probability for
selecting a jet at angle $\theta$ if it has Lorentz factor $\gamma_f$.
However, given that we observe $\beta_\mathrm{app}$ and not $\gamma$,
we must consider also the probability $p(\theta|\beta_\mathrm{app,f})$;
i.e., the probability of finding a jet at the angle $\theta$ if
it has a fixed $\beta_\mathrm{app}$.  We again use a slice of the
Monte--Carlo simulation, now for $14.5<\beta_\mathrm{app}<15.5$, to get
the probabilities shown in Figure~\ref{f:beam_par}d. The probability
density curve is broad, and as $\beta_\mathrm{app}$ decreases it becomes
more peaked.  The median value of $\beta_\mathrm{app}\sin\theta$ is
shown with the dashed line in Figure~\ref{f:prob}b, as a function
of $\beta_\mathrm{app}$.

\begin{figure}[t] 
\centering
\resizebox{1.00\hsize}{!}
{
\includegraphics[angle=0,trim=0.5cm 10cm 6.5cm 0.8cm]{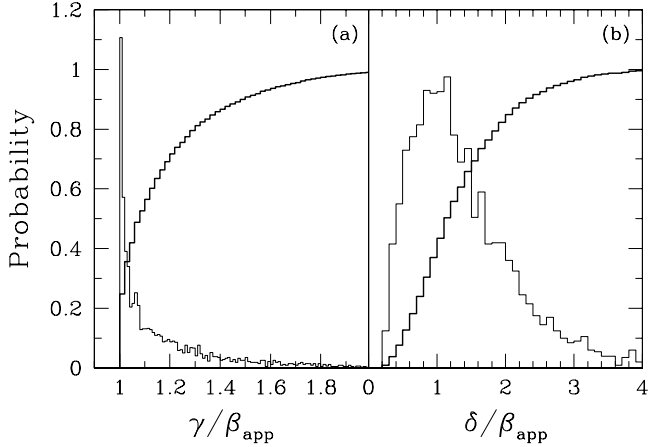}
}
\caption{\label{f:prob_curves}
Probability density and cumulative probability (heavy line) when
$\beta_\mathrm{app}\approx 15$.  (a) $p(\gamma|\beta_\mathrm{app,f})$;
(b) $p(\delta|\beta_\mathrm{app,f})$.
}
\end{figure}

The probability $p(\gamma|\beta_\mathrm{app,f})$ is also
of interest.  Figure~\ref{f:prob_curves}a shows an example,
for $\beta_\mathrm{app}\approx 15$. The probability is sharply
peaked at $\gamma\sim\beta_\mathrm{app}$. The median value
is $\gamma_\mathrm{med}/\beta_\mathrm{app}= 1.08$, and it
changes with $\beta_\mathrm{app}$ as shown with the solid line in
Figure~\ref{f:prob}b. The sharp peak can be understood in geometric
terms. In Figure~\ref{f:beam_par}b one sees that there is a large
range of $\theta$ over which $\gamma$ changes little from its minimum
value near $\beta_\mathrm{app}$, and Figure~\ref{f:beam_par}d shows
that most of the probability is in this range. For about half the
sources with $\beta_\mathrm{app}\approx 15$, $\gamma$ is between 15
and 16, but the other half is distributed to $\gamma=32$, as shown in
Figure~\ref{f:prob_curves}a. For lack of better information, it often
is assumed in the literature that $\gamma\approx\beta_\mathrm{app}$,
but this is not always valid.

Figure~\ref{f:prob_curves}b shows $p(\delta|\beta_\mathrm{app,f})$ and
$P(\delta|\beta_\mathrm{app,f})$ for $\beta_\mathrm{app}\approx 15$. The
curves change slowly for $\beta_\mathrm{app}^2\gg 1$. Unlike the Lorentz
factor, the probability for the Doppler factor does not have a sharp
peak. Consequently $L_o$, which varies as $\delta^2$, is poorly
constrained by $\beta_\mathrm{app}$.

In this paper a particular Monte--Carlo simulation is used, to show
probability curves in Figures~\ref{f:beam_par} and \ref{f:prob_curves},
and, numerically, to find the 4\%, 50\% and 96\% levels of the
cumulative probability distributions. These are fairly robust with
regard to evolution and parent luminosity functions.  We have compared
them among several of the simulations calculated by M.~Lister et al.,
and the variations are not enough to materially affect any of the
conclusions in this paper.

Figures~\ref{f:beam_par}--\ref{f:prob_curves} are not valid in the
non--relativistic case, where $\beta^2\ll 1$, $\gamma\approx 1$,
$\delta\approx 1$, $\beta_\mathrm{app}\approx \beta\sin\theta$, and
$p(\theta) \sim\sin\theta$. Our discussion is also not valid for 
samples selected on the basis of non--beamed emission.

\section {The Inversion Problem
}
\label{s:inversion}

VLBA observations can directly give apparent speed $\beta_\mathrm{app}$
and apparent luminosity $L$, but the Lorentz factor $\gamma$ and the
intrinsic luminosity $L_o$ are more useful.  We refer to the estimation
of the latter from the former as the inversion problem.

\begin{figure}[t]
\centering
\resizebox{1.00\hsize}{!}
{
\includegraphics[angle=0,trim=0.8cm 0cm 0.3cm 0.7cm]{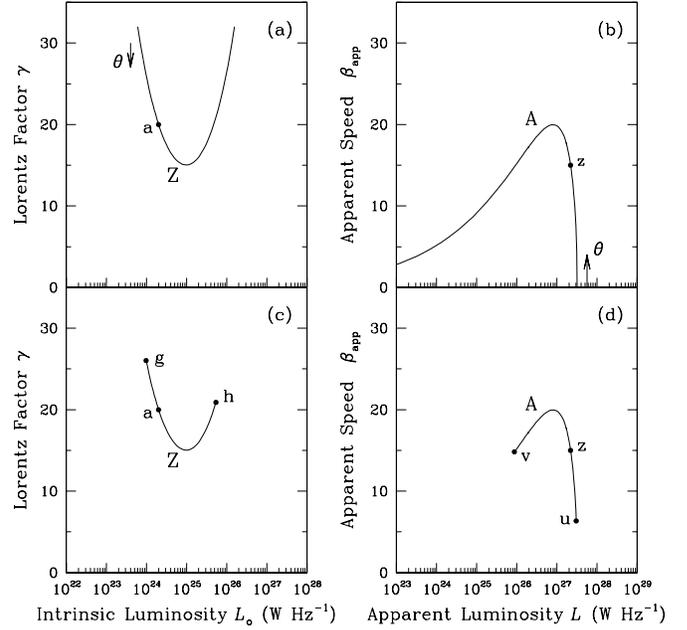}
}
\caption{\label{f:beam_curves}
Illustrating the intrinsic (left) and the observation (right)
planes for relativistic beams.  The origin point a in panel (a),
with $\gamma=20$, can be observed anywhere on the aspect curve A in
panel (b), by varying $\theta$.  The observed point $z$ in panel (b),
with $\beta_\mathrm{app}=15$, can be expressed from any point on the
origin curve Z in panel (a).  Both curves are parametric in $\theta$
with $\theta$ increasing as shown.  The maximum of the aspect curve
in panel (b) is at $\theta= 2.9\deg$ and $\beta_\mathrm{app}=19.97$.
The minimum of the origin curve in panel (a) is at $\theta=3.8\deg$
and $\gamma=15.03$.  Panel (c): as in panel (a) but with the origin curve
truncated at points $g$ and $h$, the 4\% and 96\% cumulative probability
limits, respectively.  Panel (d): as in panel (b) but with the aspect
curve truncated at points $u$ and $v$, the 4\% and 96\% cumulative
probability limits, respectively.  
} 
\end{figure}

The inversion is illustrated with Figure~\ref{f:beam_curves}. On the
left is the intrinsic plane, with axes $\gamma$ and $L_o$, and on the
right is the observation plane, with axes $\beta_\mathrm{app}$ and
$L$. Consider a source at point $a$ in Figure~\ref{f:beam_curves}a,
with $\gamma=20$ and $L_o=2\times 10^{24}$\,W\,Hz$^{-1}$.  Let it
be observed at $\theta=1.3\deg$, so that $\beta_\mathrm{app}=15.0$
and $L=2.2\times 10^{27}$~W\,Hz$^{-1}$.  This is the point $z$ in
Figure~\ref{f:beam_curves}b. Now let $\theta$ vary, and the observables
for source $a$ will follow curve A. We call A an {\em aspect} curve. It
shows all possible observable $(\beta_\mathrm{app},L)$ pairs for the
given source $a$.  The aspect curve is parametric in $\theta$, with
$\theta=0$ on the right, as shown.  The height of the curve is fixed
by the value of $\gamma$, and the location on the x-axis is fixed by
$\gamma$ and $L_o$. The width of the peak is controlled by the exponent
$n$ in equation~\ref{eq:lum}, as discussed in \S\ref{s:boost_exp}.

Now consider a source with observational parameters at point $z$
in Figure~\ref{f:beam_curves}b. What can be said about the intrinsic
parameters for this source? From equations~\ref{eq:delta}--\ref{eq:lum},
curve Z in Figure~\ref{f:beam_curves}a can be drawn; Z contains
all possible pairs of intrinsic parameters from which source $z$ can be
expressed.  We call Z an {\em origin} curve. It is parametric in $\theta$,
with $\theta=0$ on the left as shown. The curve has been truncated at
$\gamma=32$, because this is the approximate upper limit of $\gamma$
for our data, as shown in \S\ref{s:Lorentz}.

Given the lack of a constraint on $\theta$, the inversion for the
observed point $z$ in Figure~\ref{f:beam_curves}b is not unique.
Any point on the origin curve Z in Figure~\ref{f:beam_curves}a could
be its counterpart. This gives limits to $\gamma$ and $L_o$, but
they usually are broad. The limits get tighter when the probability
of observing a boosted source is considered, as in the next section.
More general results apply in a statistical sense when a sample of
sources is considered.

\subsection{Probability Cutoffs}
\label{s:prob_cutoffs}

The probabilities associated with observing beamed sources were
discussed in \S\ref{s:prob}. We now use the 4\% and 96\% cumulative probability
levels to define the regions where most of the sources will lie.
Figures~\ref{f:beam_curves}c and \ref{f:beam_curves}d repeat
Figures~\ref{f:beam_curves}a and \ref{f:beam_curves}b with the origin
curve truncated at $P(\gamma|\beta_\mathrm{app,f})=4$\% and 96\%, and
the aspect curve similarly truncated at $P(\theta|\gamma_\mathrm{f})=4$\%
and 96\%.  Note that points $g$ and $h$ do not correspond to points $u$
and $v$.  The probabilities can be seen in Figures~\ref{f:beam_par}d
and \ref{f:beam_par}c, respectively, as functions of $\sin\theta$.

The luminosities are double--valued in Figure~\ref{f:beam_curves}. The
probability cutoffs are found by integrating along curves A and Z, and
not by accumulating values of $\gamma$ or $\beta_\mathrm{app}$ along
both sides of the minimum, or peak. An example of accumulating $\gamma$
on both sides of the minimum of an origin curve is in
Figure~\ref{f:prob_curves}a.

\begin{figure}[t]
\centering
\resizebox{1.00\hsize}{!}
{
\includegraphics[angle=0,scale=0.75,trim=0.2cm 9.5cm 1.1cm 0.6cm]{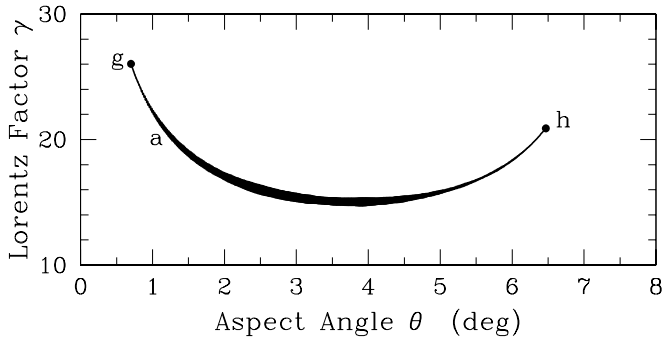} 
}
\caption{\label{f:prob_dens}
Curve Z from Figure~\ref{f:beam_curves}c is shown on the
($\gamma,\theta$) plane.  The probability density for finding a source
with $\beta_\mathrm{app}\approx 15$ is indicated by the width of the line.
Points $g$ and $h$ are the locations where the cumulative probability
$P(\theta|\beta_\mathrm{app,f})$ reaches 4\% and 96\%, respectively.
For a range of $\theta$ around the maximum probability at $\theta\sim
2.5\deg$, the value of $\gamma$ changes slowly. As shown more directly
in Figure~\ref{f:prob_curves}a, approximately half the sources with
$\beta_\mathrm{app}=15$ will have $\gamma$ between 15 and 16.  However,
$\theta$ is not similarly constrained.
}
\end{figure}

In Figure~\ref{f:beam_curves}c the points on Z have different Lorentz
factors, but all have $\beta_\mathrm{app} = 15$. The run of $\gamma$
vs $\theta$ along curve Z is shown in Figure~\ref{f:prob_dens}, which
essentially is a section of the curve in Figure~\ref{f:beam_par}b. The
probability $p(\theta|\beta_\mathrm{app,f})$, shown in
Figure~\ref{f:beam_par}d, varies slowly along this curve, and is indicated
with the line width.

From Figure~\ref{f:beam_curves}c we now have probabilistic limits
for the intrinsic parameters of the observed source $z$. Points $g$,
$h$, and the minimum give $15<\gamma<25.6$ and $1.0\times 10^{24}$
W Hz$^{-1} <L_o<5.4\times10^{25}$ W Hz$^{-1}$.  Note that these values
do not describe a closed box on the $(\gamma, L_o)$ plane. Rather, the
possible values must lie on curve Z. The highest $\gamma$ goes with the
lowest intrinsic luminosity, the lowest $\gamma$ goes with an intermediate
luminosity, and the highest luminosity goes with an intermediate $\gamma$.

A large survey will likely contain other sources with $\beta_\mathrm{app}$
near that of source $z$. They will have various luminosities and
will form a horizontal band in Figure~\ref{f:beam_curves}d. That
group of sources will have a distribution of $\gamma$ with
minimum $\gamma_\mathrm{min}\approx \beta_\mathrm{app}$ and median
$\gamma_\mathrm{med} \approx 1.1~\beta_\mathrm{app}$, according to
Figure~\ref{f:prob}b. This means that, for any individual
source, it is reasonable to guess that $\gamma$ is a little larger
than $\beta_\mathrm{app}$, although that guess will be far off for
some of the objects. It is correct to say that about half the survey
sources with $\beta_\mathrm{app}\approx 15$ will have $15<\gamma<16$,
and that about 95\% of them will have $15<\gamma<25.6$. The value 95\%
results from the 4\% above point $g$ in Figure~\ref{f:beam_curves}c,
and 1\% above $\gamma=25.6$ when the curve is continued above point $h$.

\section {The Data
}
\label{s:data}

The 2--cm VLBA survey consisted of repeated observations of 225 compact
radio sources, over the period 1994--2002.  Since that
time the MOJAVE program \citep{LH05} has continued observing a smaller
but statistically complete sample of AGN.  Most of the sources have a
``core--jet" structure, with a compact flat--spectrum core at one end of
a jet, and with less--compact features moving outward, along the jet.
The VLBA images were used to find the centroids of the core and the
components, at each epoch, and a least--squares linear fit was made
to the locations of the centroids, relative to the core. The apparent
transverse velocity was calculated from the angular velocity and the
redshift.  See \citetalias{Kellermann_etal04} and E. Ros et al. 2007,
in preparation, for details.

Each component speed is assigned a quality factor Excellent, Good, Fair,
or Poor according to criteria presented in
\citetalias{Kellermann_etal04}, but only the 127 sources with E or G
components are used here. Eight of the 127 are conservatively classified
by us as Gigahertz--Peaked--Spectrum (GPS) sources. This classification
is given only to sources that have always met the GPS spectral criteria
given by \cite{DBO97}, and is based on RATAN monitoring of broad--band
instantaneous radio spectra of AGN \citep{KNK99}\footnote{See also
spectra shown on our web site
{\sf{}\url{http://www.physics.purdue.edu/astro/MOJAVE/}}}.  In GPS
sources the bulk of the radiation is not highly beamed, as it must be if
our model is to be applicable, and we omit the GPS sources from this
study. The final sample contains 119 sources, comprising 10 galaxies, 17
BL~Lac objects, and 92 quasars, as classified by \cite{VCV03}. (See
classification discussion in \citetalias{Ketal05}.)  The sample and the
$(\beta_\mathrm{app},L)$ values used here are given on our web
site$^10$. The $\beta_\mathrm{app}$ data are updated from values in
Paper III with the addition of results from more recent epochs given in
E. Ros et al. 2007, in preparation, and on the web site.


\begin{figure*}[t]
\centering
\resizebox{0.81\hsize}{!}
{
\includegraphics[angle=0,trim=0cm 0cm 0cm 0.8cm]{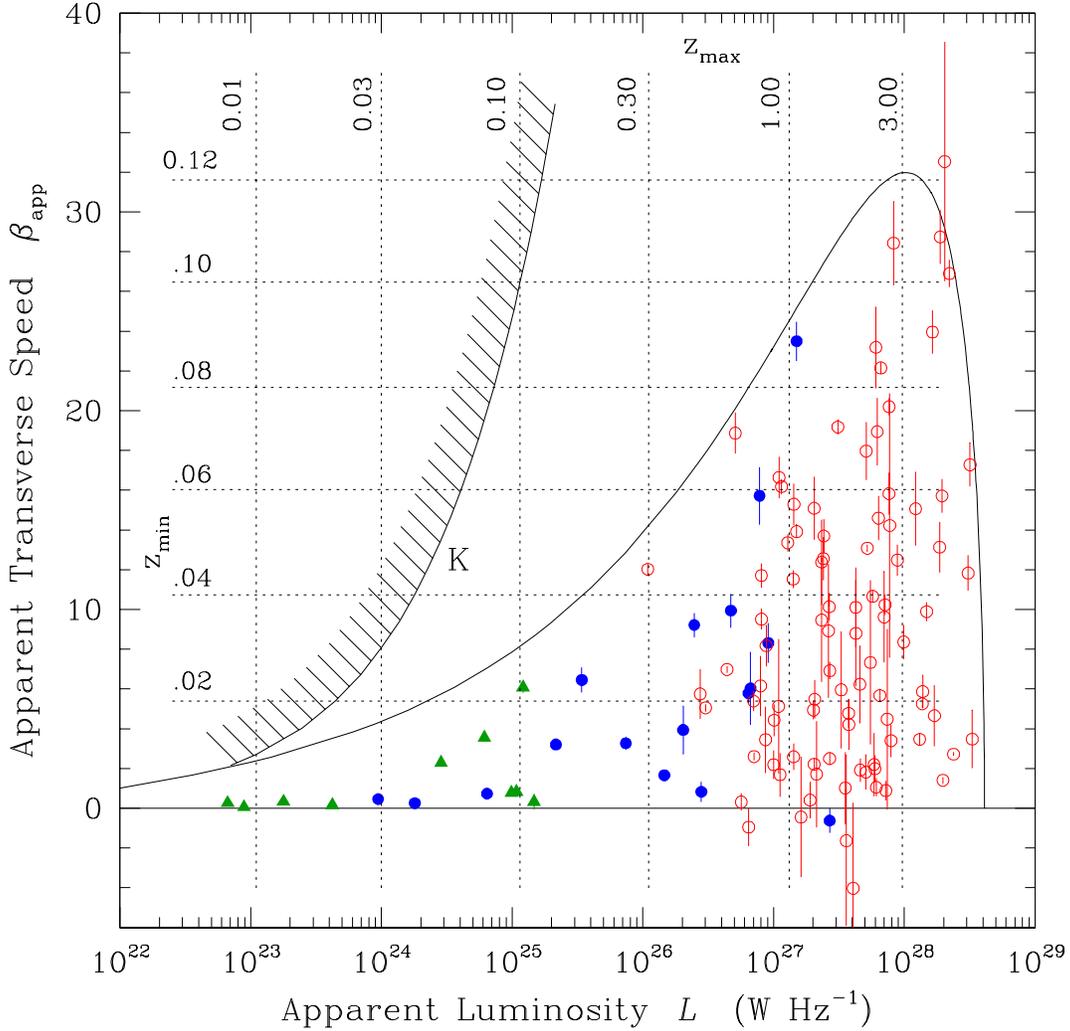}
}
\caption{\label{f:beta_L}
Values of apparent transverse speed, $\beta_\mathrm{app}$, and apparent
luminosity, $L$, are plotted for the fastest E or G component in 119
sources in the 2--cm VLBA survey. The aspect curve is the locus of
($\beta_\mathrm{app}, L$) for sources with $\gamma=32$ and $L_o=1\times
10^{25}$\,W\,Hz$^{-1}$, as $\theta$ varies. Curve K is an observational
limit set at $S_\mathrm{VLBA,med}=0.5$\,Jy and $\mu=4$\,mas\,yr$^{-1}$;
the hatched region is usually inaccessible. The horizontal lines are
the minimum values of redshift, $z_\mathrm{min}(\beta_\mathrm{app})$,
for which the angular velocity is below the limit,
$\mu<4$~mas~yr$^{-1}$. The vertical lines are the maximum values of
redshift, $z_\mathrm{max}(L)$, for which the flux density is above the
limit, $S>0.5$~Jy. See \S\ref{s:selection_effects}. Red open circles
are quasars; blue full circles, BL~Lacs; green triangles, galaxies.
}
\end{figure*}

Values of ($\beta_\mathrm{app}, L$) for the 119 sources are plotted in
Figure~\ref{f:beta_L}.  Error bars are derived from the least--squares
fitting routine for the angular velocity.  The luminosities are
calculated, for each source, from the median value of the ``total" VLBA
flux densities, over all epochs, as defined in \citetalias{Ketal05}.
$S_\mathrm{VLBA,med}$ is the integrated flux density seen by the VLBA,
or the fringe visibility amplitude on the shortest VLBA baselines.
The luminosity calculation assumes isotropic radiation. Error bars are
not shown for the luminosities. Actual errors in the measurement of
flux density are no more than 5\% \citepalias{Ketal05}, but most of the
sources are variable over time (see \citetalias{Ketal05}, Figure~11).



An aspect curve for $\gamma=32$, $L_o= 10^{25}$\,W\,Hz$^{-1}$ is shown
in Figure~\ref{f:beta_L}. It forms a close envelope to the data points
for $L>10^{26}$\,W\,Hz$^{-1}$. At lower luminosity the curve is well
above the data, and, as shown in \S\ref{s:distros} and \S\ref{s:QB},
lower aspect curves should be used there to form an envelope. A plot
similar to the one in Figure~\ref{f:beta_L} is in \cite{V95}, for the
early data from the Caltech--Jodrell Bank 6--cm survey \citep{TAY96}.
Although no aspect curve is shown in \cite{V95}, it is clear that the
general shape of the distribution is similar at 6 and 2~cm. The
parameters of the aspect curve in Figure~\ref{f:beta_L} are used in
\S\ref{s:distros} to derive limits to the distributions of $\gamma$ and
$L_o$ for the quasars.

\subsection{Selection effects}
\label{s:selection_effects}

A striking feature of Figure~\ref{f:beta_L} is the lack of sources to the
left of the aspect curve; i.e., we found no high--$\beta_\mathrm{app}$,
low--$L$ sources.  We recognize two possible selection effects which
might influence this, the lower flux density limit to the survey, and
the maximum angular velocity we can detect.  We now combine these to
derive a limit curve.

The 2--cm survey includes sources stronger than 1.5 Jy for northern
sources, and 2.0 Jy for southern sources \citepalias{kik98}. Additional
sources which did not meet these criteria, but were of special
interest, are also included in the full sample. However, here we are
using the median VLBA flux density values from \citetalias{Ketal05}
for the sub--sample of 119 sources for which we have good quality
kinematic data, and the median of these values is 1.3 Jy.  We choose
$S_\mathrm{min}=0.5$~Jy as the lower level of ``detectability,''
although 10\% of the sources are below this limit. The completeness
level actually is higher, probably close to 1.5~Jy, but the survey
sources form a representative sample of the population of sources with
$S_\mathrm{VLBA,med}>0.5$~Jy.

The angular velocity limit, $\mu_\mathrm{max}$, is set by a number of
factors, including the complexity and rate of change of the brightness
distribution, the fading rate of the moving components, and the interval
between observing sessions. These vary widely among the sources, and
there is no easily quantified value for $\mu_\mathrm{max}$. In practice,
we adjusted the observing intervals for each source according to these
factors, with $\Delta T$ being about one year in most cases. This was
usually sufficient to eliminate any ambiguity in defining the angular
velocity as seen on the ``speed plots,'' the position vs.\ time plots
shown in Figure~1 of \citetalias{Kellermann_etal04}. For some sources
there was little or no change in one year, and these were then observed
less frequently. For others, a one year separation was clearly too long,
and they were observed more frequently, typically twice per year for
complex sources. The fastest angular speeds we measured were
$\lesssim 2$ milliarcsec (mas) yr$^{-1}$, and we saw no evidence for
faster motions that would require more frequent observations. It is
important to note that even programs with shorter sampling intervals,
down to every 1 or 2 months, have not detected many speeds over 1 mas
yr$^{-1}$, and none significantly larger than 2 mas yr$^{-1}$
\citep{Gom01, Hom01, J05}.

A rough limit on our ability to identify very fast components is given
by our typical one year observing interval and the fading behavior of
jet components.  From an analysis of six sources, \cite{Hom02} found
that the flux density of jet components fades with distance from the core
as $R^{-1.3}$.  If a jet component is first identified at a separation
of 0.5 mas with a flux density of 50~mJy, that component will probably
have faded from view when it is 4 or 5 mas away, where it will have a
flux density of only a few mJy. Such a component, appearing just after a
set of observations, could fade from view before the next observation a
year later, if it was moving at $\gtrsim 4$ mas yr$^{-1}$. In practice,
however, we would be likely to observe such a source in the middle of
its cycle, and it would appear to have jet components a few mas from the
core which flicker on and off in an unpredictable fashion.  So while we
would not have been able to measure the actual speed of such a source,
it would have been identified in our sample as unusual, and followed--up
with more frequent observations. Given that we identified no such objects,
we take 4 mas yr$^{-1}$ as a reasonable upper limit to the speeds we
are sensitive to with our program.

\begin{figure*}[t]
\centering
\resizebox{0.81\hsize}{!}
{
\includegraphics[angle=0,trim=0cm 0cm 0cm 0.8cm]{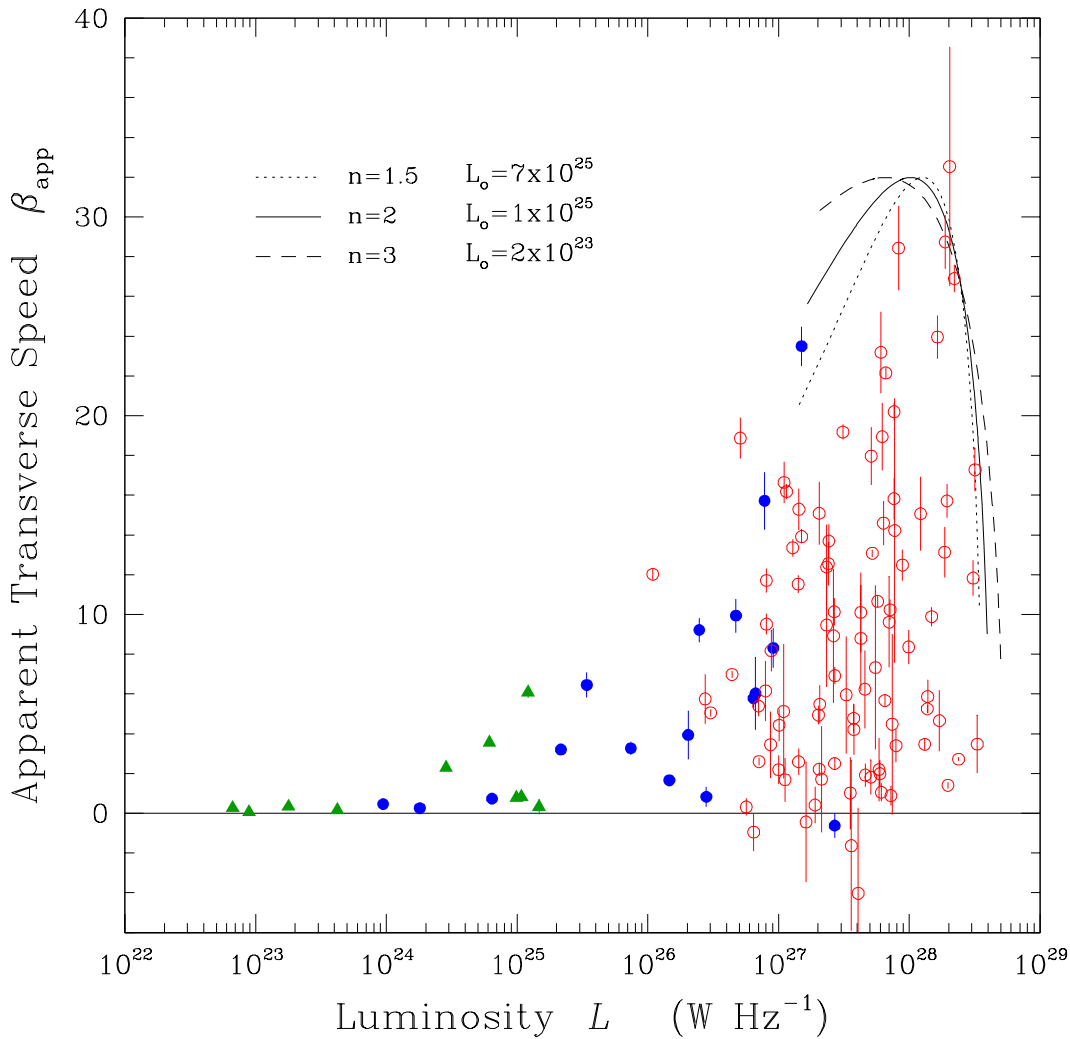}
}
\caption{\label{f:width_n}
As in Figure~\ref{f:beta_L}
but with curves for 3 values of n, the Doppler boost
exponent. The curves all have $\gamma=32$, and are truncated at the 4\%
and 96\% cumulative probability limits. Values of $L_o$ are adjusted
to optimize the fit near the top and the right--hand side.
}
\end{figure*}

It is possible that some components could fade more rapidly than the above
estimate, and if so, our limit would have to be reduced accordingly.
There is some evidence that rapid fading occurs at 43~GHz, and in
\S\ref{s:QB} we describe a source with a component moving more rapidly
at 43~GHz than at 15~GHz.  It is likely that the difference is due
to a combination of a fast fading rate and better angular resolution,
combined with the shorter observing intervals, at 43~GHz. Even here,
however, the observed speed at 43~GHz is well under the limit.

Curve K in Figure~\ref{f:beta_L}, parametric in redshift, is calculated
from the limits $S=0.5$~Jy and $\mu=4.0$~mas~yr$^{-1}$.  The hatched
region to the left of the curve is inaccessible to our observations except
in special circumstances, such as when the brightness distribution is
simple and there is only one feature in the jet.  The horizontal lines in
Figure~\ref{f:beta_L} show the minimum redshift associated with a value of
$\beta_\mathrm{app}$, set by the distance at which $\mu=4$~mas~yr$^{-1}$;
while the vertical lines show the maximum redshift associated with a
value of luminosity, set by the distance beyond which the flux density is
below 0.5~Jy. Thus, every point to the right of curve K has a range of
redshift within which it is observable, and that range fixes a spatial
volume.  Inspection of the diagram shows that the volume goes to zero
at the limit curve and increases towards the envelope. This gradient
constitutes the selection effect. Sources are unlikely to be found near
the limit curve because the available volume is small. The volume inreases
towards the envelope; and, for example, at $L=10^{26}$~W\,Hz$^{-1}$,
$\beta_\mathrm{app}=20$, the range $0.08\lesssim z\lesssim 0.30$ is
available. In the sample of 119 sources we use, there are 10 sources in
this range, all of them, evidently, far from the region in question. At
$L=10^{25}$~W\,Hz$^{-1}$, $\beta_\mathrm{app}=10$, the range $0.04\lesssim
z\lesssim 0.10$ is available and 6 of the survey sources are in this
range; again, none of them is near the region in question.  Hence, the
lack of observed sources to the left of the envelope is not a selection
effect; but rather, must be intrinsic to the objects themselves.


\subsection{The Fast Sources}
\label{s:fast_sources}

The four sources we found with $\mu\ge1$\,mas\,yr$^{-1}$ are all
in the \citetalias{VC94} compilation. \citetalias{VC94} listed four
additional sources with $\mu\ge 1$\,mas\,yr$^{-1}$: \objectname{M87},
which has a fast long--wavelength (18~cm) component far from the
core, \objectname{Cen~A}, which is in the southern sky and therefore
not included in our study, and two others, \objectname{Mrk~421}
and \objectname{1156+295}, where our measured values are well below
1\,mas\,yr$^{-1}$ \citepalias{Kellermann_etal04}.

We note that, with years of increasingly better observations on
more objects, the known number of sources with fast components
has not increased. There are only 5 compact jets that show
$\mu>1$\,mas\,yr$^{-1}$ at 15~GHz, within our flux density range.  These
are all nearby objects and include three galaxies, \objectname{3C~111},
\objectname{3C~120}, and \objectname{Cen~A}; one BL~Lac object,
\objectname{BL~Lac} itself; and one quasar, \objectname{3C~273}.
Monthly monitoring at higher resolution by \citetalias{J05} detected
5 sources (out of 15) that had $\mu>1$\,mas\,yr$^{-1}$. We found 4
of these, but we measured $\mu<1$\,mas\,yr$^{-1}$ for their fifth
object, \objectname{1510$-$089}. In addition, they measured $\mu\sim
1$\,mas\,yr$^{-1}$ for \objectname{0219+428} (\objectname{3C~66A}),
but it has low flux density and is not in our survey.

\subsection{The Boost Exponent}
\label{s:boost_exp}

The exponent $n$ in equation~\ref{eq:lum} controls the
sharpness of the peak of the aspect curve. Figure~\ref{f:width_n} shows
the data with three aspect curves for $\gamma=32$, with different
values of $n$. The curves have been truncated at the 4\% and 96\%
probability limits, and the values of $L_o$ have been adjusted so
that the curves roughly match the right-hand side of the data. The
probability was calculated with equation (A15) from \citetalias{VC94},
as the simulation described in Appendix~\ref{appendix} uses $n=2$ and
has not been calculated for other values of $n$.

It is important to compare the curves with the data only in the region
where the probability is significant. From Figure~\ref{f:width_n} there
is no strong reason to pick one value of $n$ over another.  However,
if $n=3$, the boosting becomes so strong that strong distant quasars,
near the peak of the distribution, have $L_o$ as small as the jets in
weak nearby galaxies, which (we argue in \S\ref{s:galaxies}) are only
mildly relativistic.  This is unrealistic, and we conclude that $n <
3$. The value $n=2$ has a theoretical basis (\S\ref{s:assum}) and we
have adopted it here. Note the large range in intrinsic luminosity
corresponding to different values of $n$.  Since $n$ is not known with
precision, the intrinsic luminosities have a corresponding uncertainty.

\section {The Peak Lorentz Factor
}
\label{s:Lorentz}

Beaming is a powerful relativistic effect that supplies a strong
selection mechanism in high--frequency observations of AGN.  Consider a
sample of randomly oriented, relativistically boosted sources that have
distributions in redshift, intrinsic luminosity, and Lorentz factor.
Make a flux density--limited survey of this sample. \citetalias{VC94}
and \citetalias{LM97} have shown that in this case the selected sources
will have a maximum value of $\beta_\mathrm{app}$ which closely approaches
the upper limit of the $\gamma$ distribution, even for rather small sample
size.  This comes about because the probability of selecting a source is
maximized near $\theta=0.6\theta_c$ where $\beta_\mathrm{app}\approx
0.9\gamma$ (see Figure~\ref{f:beam_par}c); and there is a high
probability that, in a group of sources, some will be at angles close
to $\theta_\mathrm{c}$, where $\beta_\mathrm{app}\approx\gamma$. Hence,
because $\beta_\mathrm{app,max}\approx 32$, the upper limit of the
$\gamma$--distribution is $\gamma_\mathrm{app,max}\approx 32$.

\section {The Distributions of $\gamma$ and $L_o$
}
\label{s:distros}


We showed in \S\ref{s:selection_effects} that the lack of sources to the
left of the envelope is not a selection effect, but is intrinsic to the
objects. Since the envelope is narrow at the top, $\beta_\mathrm{app}$
and $L$ are correlated; high $\beta_\mathrm{app}$ is found only in sources
that also have high $L$, but low $\beta_\mathrm{app}$ is found in sources
with all values of $L$. This translates into $\gamma$ having a similar
correlation with $L_o$, for the quasars. The $\gamma$ distribution will be
similar to the $\beta_\mathrm{app}$ distribution in Figure 7, but flatter;
with many points shifted up, but nearly all by less than a factor 2 above
$\beta_\mathrm{app}$.  The $L_o$ distribution will remain more spread
out at low $\gamma$ than at high $\gamma$, leading to the correlation
that the highest $\gamma$ are found only in jets with high intrinsic
luminosity.  This is consistent with a result from \citetalias{LM97},
that Monte--Carlo simulations with negative correlation between $\gamma$
and $L_o$ give a poor fit to the statistics of the flux densities from
the Caltech--Jodrell Bank survey \citep{TAY96}.

The good fit of an aspect curve as an envelope to the data in
Figure~\ref{f:beta_L} suggests that the parameters of the curve,
$\gamma=32$ and $L_o= 10^{25}$\,W\,Hz$^{-1}$, reflect the peak values
of $\gamma$ and $L_o$ in the population.  The distribution of $\gamma$
may be a power law, as suggested by \citetalias{LM97} and, as discussed
in \S\ref{s:Lorentz}, $\gamma_\mathrm{max}=32$ is close to the maximum
value in the distribution. We now consider constraints on the peak value
for $L_o$.

\begin{figure*}[t]
\centering
\resizebox{0.81\hsize}{!}
{
\includegraphics[angle=0,trim=0cm 0cm 0cm 0.8cm]{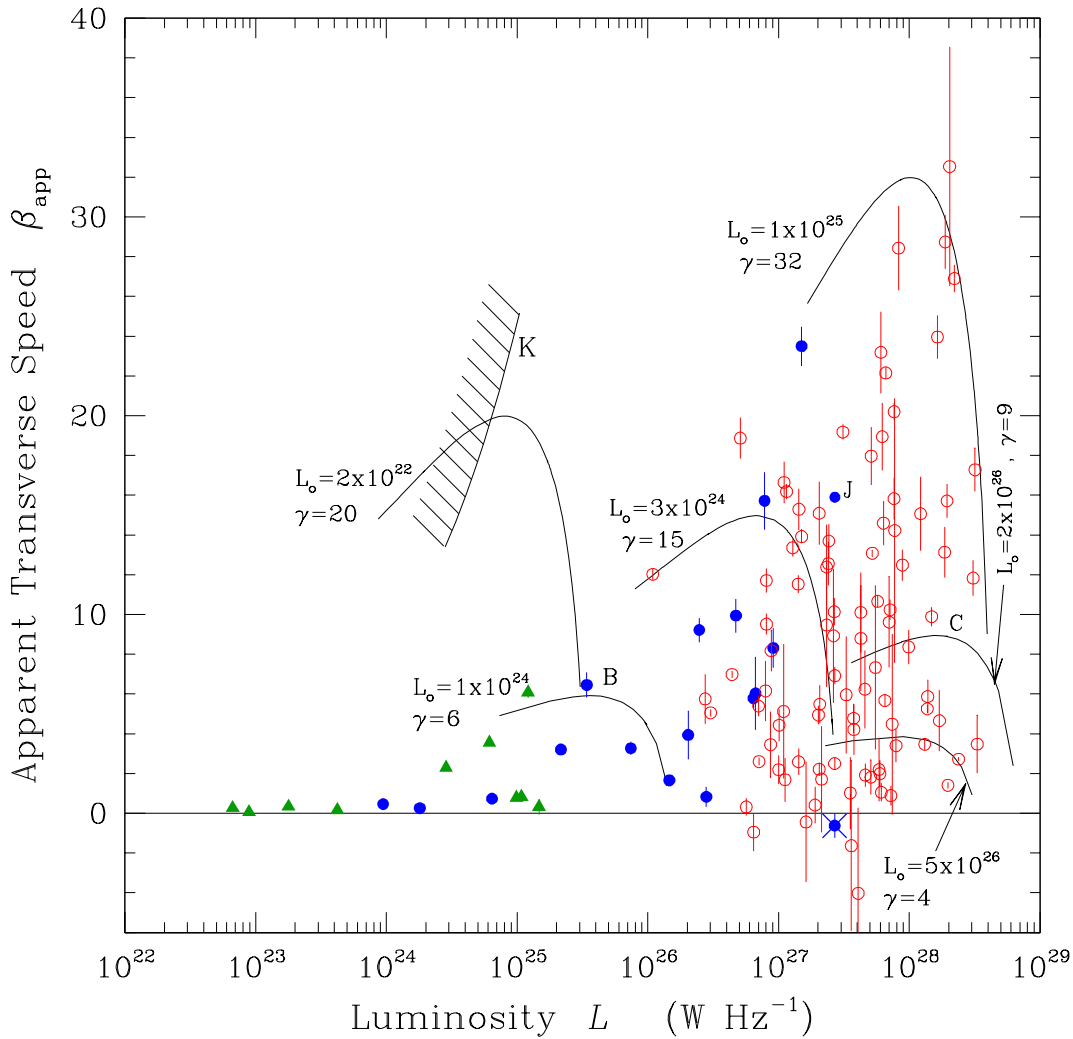}
}
\caption{\label{f:apparent_gamma}
The data are plotted as in Figure~\ref{f:beta_L}. The aspect curves
are truncated at the 4\% and 96\% probability levels.  At the peak of
each curve, $P(\theta)\approx 0.75$; i.e. about 3/4 of the probability
of selecting a source with this value of $\gamma$ and $L_o$ is on the
right side of the curve.  The cross close to $\beta_\mathrm{app}=0$
marks the source 1803+784, and point J is the same source but
with the $\beta_\mathrm{app}$ value at 43~GHz from \citetalias{J05},
see text.  Curve K is a short section of the limit curve K from
Figure~\ref{f:beta_L}.
}
\end{figure*}

Figure~\ref{f:apparent_gamma} is similar to Figure~\ref{f:beta_L},
but with several aspect curves, each showing only the region
$0.04<P(\theta|\gamma_\mathrm{f})<0.96$. The envelope is now formed
by a series of aspect curves, with successively lower values of
$\gamma$. Most of the sources will have $\gamma$ rather close to
$\beta_\mathrm{app}$, but some will have $\gamma$ substantially
greater (see Figures~\ref{f:prob}b and \ref{f:prob_curves}a).
In Figure~\ref{f:apparent_gamma} these latter sources will not lie near
the top of an aspect curve, but will be down from the peak.  It is more
likely that they will be at small angles ($\theta<\theta_\mathrm{c}$)
than at large angles.

Consider the \objectname{BL~Lac} marked B, near the intersection of curves
$\gamma=6$ and $\gamma=20$. It could be on either curve, but it is near
the low--probability region of curve $\gamma=20$.  For every source on
curve $\gamma=20$ near the intersection, there should be several farther
up the curve. Note that the $\gamma=20$ curve intersects the limit curve K
to the left of the peak. There is little available redshift volume at the
peak, but the volume increases rapidly at lower $\beta_\mathrm{app}$,
and the lack of sources there means that source B is unlikely to
have $\gamma=20$.  Alternatively, it could be close to an extension of
curve $\gamma=15$, but then it is again in a low--probability region.
There are a number of sources near the peak of curve $\gamma=15$, and
B could be a high--angle version of one of them.  But the probability
of that is well below 0.04, and there can be few such sources in the
entire sample of 119.  We conclude that the galaxies and BL~Lacs on
the left side of the distribution ($L<3\times 10^{25}$\,W\,Hz$^{-1}$),
with high confidence, are not off-axis versions of the powerful quasars
(curves $\gamma = 15$ and 32), nor are they high--$\gamma$, low--$L_o$
sources (curve $\gamma=20$).

Source B in Figure~\ref{f:apparent_gamma} is the eponymous object
\objectname{BL~Lac}, \objectname{2200+420}. \cite{DMM00} studied
\objectname{BL~Lac} in detail, and showed that the jet lies on a helix
with axis $\theta= 9\deg$ and pitch angle $2\deg$. If $\theta=9\deg \pm
2\deg$ is combined with our value for the apparent transverse speed,
$\beta_\mathrm{app}=6.6\pm 0.6$, then $\gamma=7\pm 1$. This agrees
with our conclusion above.

Now consider the sources near point C in Figure~\ref{f:apparent_gamma}, at
$L=3\times 10^{28}$\,W\,Hz$^{-1}, \beta_\mathrm{app}=9$.  They could have
$\gamma\approx 9$, but in that case there should be several others down
the $\gamma=9$ curve to the right, where most of the probability lies.
But there are none there.  Any aspect curve with a peak farther to the
right is unlikely to represent any of the measured points, and so curve
$\gamma=9$ is about as far to the right as should be considered.  If the
sources at C are on curve $\gamma=9$, then their intrinsic luminosity is
an order of magnitude greater than that for the sources near the top of
the distribution, the fastest quasars. To avoid a negative correlation
between $\gamma$ and $L_o$, some of the sources near point C should
have $\gamma=20$ or more, with the appropriate small values of $\theta$.
However, others near the right-hand side of the distribution might well
have $\gamma\sim 9$ or smaller.  This means that the distribution of $L_o$
could extend up to $10^{26}$\,W\,Hz$^{-1}$.

\section {Quasars and BL~Lacs with $\beta_\mathrm{app}<3$
}
\label{s:QB}

Twenty-two of the 92 quasars and 3 of the 13 powerful BL~Lacs in
Figure~\ref{f:apparent_gamma} ($L>3\times 10^{25}$\,W\,Hz$^{-1}$)
have $\beta_\mathrm{app}<3$, and have low probability if $\gamma>10$.
What are the intrinsic properties of this group?  We consider three
possibilities. (i) They are high--$\gamma$ sources seen nearly end-on,
and have $P(\theta)<0.04$. We expect only a few such end-on sources
out of a group of 105. Most of the low--speed quasars cannot be
explained this way. (ii) They are low--$\gamma$ high--$L_o$ sources,
and have $\gamma\sim 3$.  We discussed this above for point C in
Figure~\ref{f:apparent_gamma} with $\beta_\mathrm{app}=9$; now we
are considering $\beta_\mathrm{app}<3$, and the argument is stronger.
Unless the most intrinsically luminous sources have low $\gamma$, this
option is not viable.  (iii) A more likely situation is that many of
these low--$\beta_\mathrm{app}$ components appear to be slow because
$\gamma_p <\gamma_b$. 


In support of comment (iii), we note that one of the slow objects,
\objectname{1803+784}, was also observed by \citetalias{J05}
at 43~GHz. They find $\beta_\mathrm{app} =15.9\pm 1.9$, whereas,
at 15~GHz, we found $\beta_\mathrm{app}=-0.6\pm 0.6$.  The higher
resolution at 43~GHz is crucial in detecting fast components in sources
like this, because they are within 1 mas of the core, at or below the
resolution limit at 15~GHz. In Figure~\ref{f:apparent_gamma} the 15~GHz
speed for \objectname{1803+784} is shown with a cross; and the 43~GHz
$\beta_\mathrm{app}$ value, with the 15~GHz luminosity, is shown with
the letter J. It is likely that we have reported a component speed that
is not indicative of the beam speed for \objectname{1803+784}.

In Figure 8, source \objectname{1803+784} is in a cluster of objects that,
formally, have negative speed. However, they all are within 1$\sigma$
of zero, and their negativity is of little significance.  A number of
other sources have components with similar formally negative speeds,
but in addition they have a component with a larger positive speed. In
this paper we have only used the fastest component in each source.

\section{Galaxies
}
\label{s:galaxies}

\begin{deluxetable*}{lllllcccr}
\tabletypesize{\small}
\tablewidth{0pt}
\tablecaption{Galaxies with $\beta_\mathrm{app} > 1$
\label{t:galaxies}
}
\tablewidth{0pt}
\tablehead{
\colhead{IAU name} & \colhead{Alias} & \colhead{Type} & \colhead{Redshift} & \colhead{$\beta_\mathrm{app}$} & \colhead{$D_\mathrm{var}$} & \colhead{$\gamma$} & \colhead{$\theta$ (deg)} & \colhead{$L_o$}\\
\colhead{(1)}      & \colhead{(2)}   & \colhead{(3)}   & \colhead{(4)}      & \colhead{(5)}                  & \colhead{(6)}              & \colhead{(7)}      & \colhead{(8)}      & \colhead{(9)}
}
\startdata
\objectname{0415+379} & \objectname{3C~111}   & Sy1 & 0.049 & 6.1$\pm$0.1 & 3.4   & 7.3 & 15 & $4.4\times 10^{23}$  \\
\objectname{0430+052} & \objectname{3C~120}   & Sy1 & 0.033 & 3.6$\pm$0.2 & 2.4   & 4.1 & 22 & $4.8\times 10^{23}$  \\
\objectname{1845+797} & \objectname{3C~390.3} & Sy1 & 0.057 & 2.3$\pm$0.1 & 0.9   & 3.8 & 42 & $3.1\times 10^{24}$ \\
\enddata
\end{deluxetable*}

The points in Figure~\ref{f:beta_L} appear to run smoothly from low to
high apparent luminosity, suggesting that the different types of objects
might be closely related.  However, the smoothness is supplied by the
BL~Lac objects, which connect the galaxies and quasars that otherwise
are widely separated in apparent luminosity. In addition, the galaxies
all have $z\le 0.2$, and nearly all the quasars have $z>0.4$. The
separation is at least partly the result of our restricted sensitivity,
coupled with the luminosity functions. We cannot observe ``galaxies''
at high redshift because our sensitivity is too low; and we see few
``quasars'' at low redshift because their local space density is so low.
In this section we consider whether the galaxies and quasars form separate
classes, or, in particular, whether the galaxies might be high--$\theta$
counterparts of the more luminous sources \citep{UP95}. 

Three galaxies have superluminal components, and their speeds place them
with the lower--speed quasars, as seen in Figure~\ref{f:apparent_gamma}.
These fast galaxies, shown in Table 1, all have broad emission
lines and are classified as Sy1; they are at low redshift and are
highly variable at radio wavelengths.  The obscuring torus paradigm
for Sy1 galaxies \citep{AM85} suggests that they are not at large
values of $\theta$, and this is confirmed by the observed values of
$\beta_\mathrm{app}$, which show that $\theta$ must be less than
$\theta_\mathrm{max}=2\arctan\beta_\mathrm{app}^{-1}\sim 20\deg$
to 45\deg. To estimate values of $\gamma, \theta$, and $L_o$
for these galaxies, we combine the measured $\beta_\mathrm{app}$
with a variability Doppler factor, $D_\mathrm{var}$, derived
from the time scale and strength of variations in flux density
\citep[e.g.,][]{Coh04}. $D_\mathrm{var}$ is given by \citetalias{J05}
for \objectname{0415+379} and \objectname{0430+052}, and by \cite{LV99}
for \objectname{1845+797}. We have converted the last value to the
cosmology used in this paper, and use an intrinsic brightness temperature
$T_\mathrm{b}=2\times 10^{11}$\,K. This is a characteristic lower
limit for sources in their highest brightness states \citep{Hom06},
and should be more appropriate than the canonical equipartition value,
for variability measurements based on flux density outbursts.  Note that
\citetalias{J05} use a different procedure to calculate $D_\mathrm{var}$,
and do not assume an intrinsic temperature. The $D_\mathrm{var}$ are
model--dependent and their reliability is difficult to assess.


The Lorentz factors for the quasars are not estimated in this paper,
but from Figure 3 it can be seen that many of them must have Lorentz
factors close to their apparent speeds.  Thus, from Table 1 and Figure
8, the Lorentz factors of the superluminal galaxies are comparable
with those for the slower quasars.  Their luminosities, however, do
not overlap with those for the quasars, indicating that they are a
different population.

Only one of the seven slow galaxies ($\rm\beta_{app}<1$) has
$\beta_\mathrm{app}$ consistent with zero (2$\sigma$). The others show
definite motions and several must be at least mildly relativistic,
with $\rm\beta_{app}>0.3$. The galaxies with $\rm\beta_{app}<0.3$
cannot contain a highly relativistic jet, for that would force
$\theta$ to be unacceptably small. For example, if $\beta_\mathrm{app}=0.3$
and $\gamma=5$, then $\theta=0.35\deg$ and $\delta=9.9$. This gives
$\theta/\theta_\mathrm{c}=0.03$, which is extremely unlikely, for
an object with $\gamma=5$ (Figure~\ref{f:beam_par}c).  In any event,
all proposed galaxy--quasar unifications place the galaxy at a high
angle, where the flux density and apparent speed are reduced, but
a relativistic beam cannot show $\rm\beta_{app} = 0.3$ at any
angle not near 0\deg ~(or 180\deg).  Hence, the galaxies are neither
low--angle nor high--angle versions of the distant quasars. However,
Cygnus A may be an exception, as discussed in the next section.



\cite{Gio01} have concluded that most radio galaxies (including FRI)
contain relativistic jets.  They assumed that all sources have jets
with intrinsic bipolar symmetry, and used the measured side--to--side
ratio with a correlation between lobe power and intrinsic core power to
obtain limits on $\beta$ and $\theta$.  Our procedure may be more robust
because each source has a measured $\beta_\mathrm{app}$, and we do not
appeal to symmetry of the lobes.

\subsection{Cygnus A}
\label{s:cyg_a}

The galaxy \objectname{Cygnus~A} $(z=0.056)$ merits special discussion.
The radio lobes are exceptionally powerful, and their luminosity is
comparable to that of the most powerful and distant radio galaxies. The
jets, however, are weak.  Optical polarization studies \citep{O97}
reveal polarized broad lines and show that \objectname{Cygnus~A} is a
modest quasar. \cite{Bar95} used the front-to-back ratio of the jets
of \objectname{Cygnus~A} at 6~cm, together with $\beta_\mathrm{app}$,
to estimate $\theta$.  We repeat their analysis with our value for
$\beta_\mathrm{app}, ~0.83 \pm 0.12$, and obtain $45\deg<\theta<70\deg$.
This agrees with other estimates of the angle, including \cite{O97}
who found $\theta>46\deg$, and \cite{VB93}, who found $\theta\sim$
50\deg to 60\deg. The combination of $\beta_\mathrm{app}$ and $\theta$
gives $1.24<\gamma<1.36$ and $0.59<\beta<0.68$. \objectname{Cygnus~A}
is mildly relativistic.

On the other hand, because the lobes in \objectname{Cygnus~A} are so
powerful, we might have expected that it would have a highly relativistic 
jet.  These contradictory ideas can be reconciled with a two--component
beam, consisting of a fast spine with a slow sheath, as suggested by
numerical simulations \citep[e.g.,][]{AGU01}. The slow beam that we
see has $S_\mathrm{slow} = 1.5$~Jy. The fast beam is at a high angle to
the LOS and not seen because it is deboosted, and it must be at least a
factor of ten weaker than the slow beam; i.e., $S_\mathrm{fast}<0.15$~Jy.
If the fast beam has a Lorentz factor of about 10, then if observed at a
small angle its flux density would be up to a few hundred Jy, far higher
than observed in any other source.  But \objectname{Cygnus~A} is much
closer than most superluminal sources; and if the nearest quasar, 3C 273,
were at the distance of \objectname{Cygnus~A}, its flux density would
be within a factor 2 of our putative value for \objectname{Cygnus~A}.
If \objectname{Cygnus~A} were at $z\sim 1$, and pointed near the LOS,
it would be a normal quasar, with radio and optical luminosity somewhat
below the median for quasars. It is exceptional only because it is
accidentally nearby.  This model solves the long--standing problem of
the strong lobes combined with the weak core.

In this model the total flux density from \objectname{Cygnus~A} varies
more slowly with $\theta$ than $\delta^2$, the commonly assumed law.
If this is correct, and if it applies generally to many other radio
sources, then it will affect the usual discussions of the unification
of radio sources by aspect.  See, e.g., \cite{CCC00}, who invoke a
two--component model in their article on unification.

\section {Discussion}
\label{s:disc}

The aspect curve in Figure~\ref{f:beta_L} is a good envelope to
the quasar data, and this suggests that the relativistic beam model
is realistic.  The slow rise and rapid fall of the curve is a direct
consequence of Doppler boosting combined with time contraction, both of
which are relativistic effects. We have used the common assumption that
the moving VLBI component is traveling with the beam; i.e., $\gamma_p
\approx \gamma_b$. But if this is incorrect, for example, if the beam
and pattern speeds are independent, or often are far apart, then it is
hard to see why an aspect curve should form an envelope. In particular,
if they are independent, then some sources should have high $\gamma_p$
with low $\gamma_b$, and this could place them to the left of the
envelope in Figure~\ref{f:beta_L}. The lack of sources there has been
emphasized earlier, and is evidence that, for the fastest components in
many sources, $\gamma_p$ and $\gamma_b$ are closely  related. ,

In about a fourth of the BL Lacs and quasars the fastest component appears
to be moving slowly $(\rm\beta_{app}<3)$. These objects are discussed in
\S\ref{s:QB}, where it is shown, using probability arguments, that most
of them cannot be high--$\gamma$ sources with a small value of $\theta$.
However, other evidence suggests strongly that most of these objects
do have a highly relativistic beam. This evidence consists of rapid
variability \citep[e.g.,][]{AAH03, T05}; high apparent brightness
temperature measured both directly with VLBI \citepalias{Ketal05}
and indirectly with interstellar scintillations \citep{Kra03, Lov03};
gamma-ray emission \citep[e.g.,][]{DG95}; and observational effects due
to differential Doppler boosting, such as the Laing--Garrington effect
\citep{LAI88}. Many of these sources must have $\gamma_p\ll\gamma_b$,
as discussed in \S\ref{s:QB}.

\section{Conclusions}
\label{s:conclusions}

1. The aspect and origin curves provide a useful way to understand
the relations between the intrinsic parameters of a relativistic
beam, Lorentz factor $\gamma$ and intrinsic luminosity $L_o$, and the
observable parameters, apparent transverse speed $\beta_\mathrm{app}$
and apparent luminosity $L$.  Limits to the intrinsic parameters, for
a given observed source, are found on its origin curve that has been
truncated by probability arguments.

2. About half the sources with $\beta_\mathrm{app}>4$, that are found
in a flux--density limited survey, will have $\gamma$ within 20\%
of $\beta_\mathrm{app}$.

3. The 2--cm VLBA survey has yielded high--quality kinematic data for
119 compact radio jets. When plotted on the observation plane, they are
bounded by an aspect curve for $\gamma=32$, $L_o= 10^{25}$~W~Hz$^{-1}$,
that forms a good envelope to the data at high luminosities. From this,
with probability arguments, we find that the peak Lorentz factor in the
sample is $\gamma\approx 32$ and the peak intrinsic luminosity is $L_o\sim
10^{26}$~W~Hz$^{-1}$.

4. There is an observed correlation between $\rm\beta_{app}$ and $L$ for
the jets in quasars: high $\rm\beta_{app}$ is found only in radio jets
with high $L$. This implies a similar correlation between $\gamma$ and
$L_o$: high $\gamma$ must preferentially exist in jets with high $L_o$.

5. The Doppler--boosting exponent $n$ for a typical source in the
survey must be less than 3, or else the highly luminous jets with
the fastest superluminal speeds will have intrinsic luminosties
comparable to the slow, nearby galaxies.

6. There are too many low-speed ($\beta_\mathrm{app}<3$) quasars in
the sample, according to probability arguments. It is likely that
some of them have pattern speeds substantially lower than their
beam speeds. 

7. The galaxies have a distribution of Lorentz factor up to
$\gamma=7$; three show superluminal motion, but most are only mildly
relativistic. They are not off--axis versions of the powerful quasars.
Cygnus A may be an exception, and we suggest that it might have a
``spine-sheath" morphology.

8. Our results strongly support the common relativistic beam model
for compact extragalactic radio jets. The pattern and beam speeds
must be approximately equal, for the fastest components in many
sources.

\acknowledgments

We are grateful to the rest of the 2--cm VLBA Survey Team, who have
contributed to the data used in this paper, and for their support and
advice. We thank Tsvi Piran and Manuel Perucho for helpful discussions,
Steven Bloom for commenting on the manuscript, and the referee for
helpful comments.  The MOJAVE project is supported under National
Science Foundation grant 0406923-AST.  \mbox{RATAN--600} observations
were supported partly by the NASA JURRISS program (W--19611) and the
Russian Foundation for Basic Research (01--02--16812, 02--02--16305,
05--02--17377).  Y.~Y.~Kovalev is a Research Fellow of the Alexander
von Humboldt Foundation.  M.~Kadler has been supported in part by a
Fellowship of the International Max Planck Research School for Radio
and Infrared Astronomy and in part by an appointment to the NASA
Postdoctoral Program at the Goddard Space Flight Center, administered
by Oak Ridge Associated Universities through a contract with NASA.
D. Homan is supported by a grant from Research Corporation.  The Very Long
Baseline Array is operated by the National Radio Astronomy Observatory,
a facility of the National Science Foundation operated under cooperative
agreement by Associated Universities, Inc.  This research has made use
of the NASA/IPAC Extragalactic Database (NED) which is operated by the
Jet Propulsion Laboratory, California Institute of Technology, under
contract with the National Aeronautics and Space Administration.

\appendix
\section{Monte--Carlo Calculations}
\label{appendix}

M.~Lister et al.~(in preparation) are performing Monte--Carlo
calculations to simulate observations being made in the MOJAVE survey
\citep{LH05}. One of their simulations that illustrates the probability
densities we need is used here. The calculation selects 100,000
sources with $S>1.5$~Jy ($S>2$~Jy in the south) from a large parent
population with a distribution of Lorentz factors having a power law
with index -1.25 and peak $\gamma=32$, and a power-law distribution of
intrinsic luminosities with index -2.73 and minimum luminosity $1\times
10^{24}$~W~Hz$^{-1}$.  The model uses an evolving luminosity function
based on a fit to the quasars in the Caltech--Jodrell Bank 6--cm survey
\citepalias{LM97}.  The calculation does not assume any correlation
between the intrinsic quantities $\gamma$ and $L_o$.

\begin{figure}[t]
\centering
\resizebox{0.81\hsize}{!}
{
\includegraphics[angle=0,trim=0cm 0cm 0cm 0.8cm]{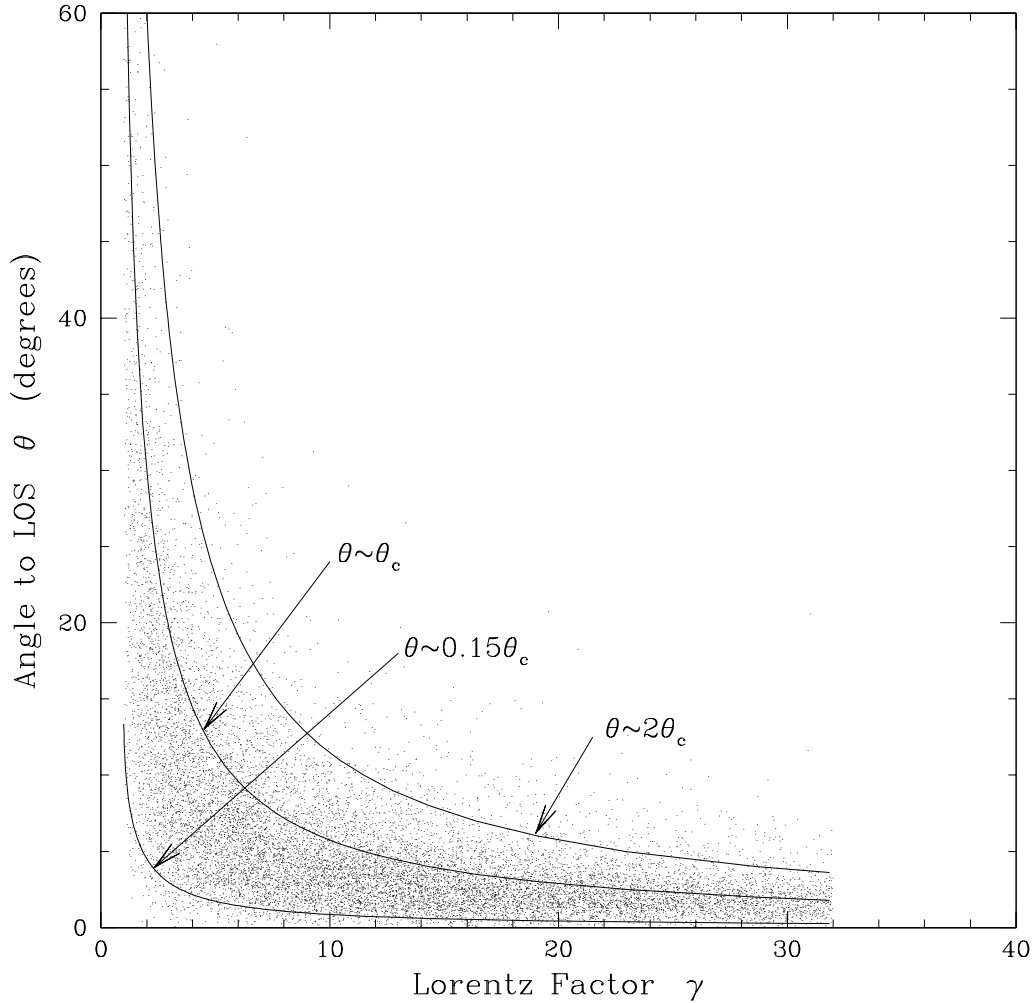}
}
\caption{\label{f:appendix}
Distribution of 14,000 sources selected randomly from the simulation. The
central line is $\gamma\sin\theta=1$, or $\theta\approx\theta_\mathrm{c}$.
The top and bottom lines are $\gamma\sin\theta = 2.0$ and
$\gamma\sin\theta = 0.15$, corresponding to $P(\theta)=$96\% and
4\%, respectively.
}
\end{figure}

Figure~\ref{f:appendix} shows 14,000 of the selected sources,
in ($\theta,\gamma$) space.  The curve $\gamma\sin\theta=1.0$
is shown and it can be seen that the majority of sources have
$\theta<\theta_\mathrm{c}$. The lines $\gamma\sin\theta=0.15$ (bottom)
and $\gamma\sin\theta=2.0$ (top) are also shown; these show the 4\% and
96\% cumulative levels for the simulation. They are used in the text as
practical limits.

Lister et al.\ discuss the distribution functions for various intrinsic
and observed quantities. Here, we only show results for subsamples
representing slices through the full distribution at constant $\gamma$,
and at constant $\beta_\mathrm{app}$. Figure~\ref{f:beam_par}c shows a
slice for $14.5<\gamma<15.5$ (N=3638); the histograms are the
probability density, $p(\theta|\gamma\approx 15)$, and the cumulative
probability, $P(\theta|\gamma\approx 15)$. The probability density
$p(\theta|\gamma_f)$ varies slowly with $\gamma$, and
Figure~\ref{f:prob}a shows $p(\sin\theta<\gamma^{-1})$, the expected
fraction of sources that will lie within their critical angle.
Figures~\ref{f:beam_par}d and \ref{f:prob}b show similar distributions
for $14.5<\beta_\mathrm{app}<15.5$ (N=3191).

\begin{thebibliography}{}

\bibitem[Agudo et al.(2001)]{AGU01}
Agudo, I., G\'omez, J.-L., Marti, J.-M., Abanez, J.-M., Marscher, A.P.,
Alberdi, A., Aloy, M.-A. \& Hardee, P. E. 2001, \apj, 549, L186

\bibitem[Aller, Aller, \& Hughes(2003)]{AAH03}
Aller, M.~F., Aller, H.~D., \& Hughes, P.~A. (2003), in ASP Conf. Ser. 
300, Radio Astronomy at the Fringe, ed.\ J.~A.\ Zensus, M.~H.\ 
Cohen, \& E.\ Ros (San Francisco: ASP), 159

\bibitem[Antonucci \& Miller(1985)]{AM85}
Antonucci, R.R.J. \& Miller, J.S. 1985, \apj, 297, 621


\bibitem[Barthel et al.(1995)]{Bar95}
Bartel, N., Sorathia, B., Bietenholtz, M.F., Carilli, C.L.,
\& Diamond, P. 1995, PNAS, 92, 11371


\bibitem[Blandford \& K{\"o}nigl(1979)]{BK79}
Blandford, R.D. \& K{\"o}nigl, A. 1979, \apj, 232, 34

\bibitem[Chiaberge et al.(2000)]{CCC00}
Chiaberge, M., Celotti, A., Capetti, A., \& Ghisellini, G. 2000,
\aap, 358, 104

\bibitem[Cohen et al.(2004)]{Coh04}
Cohen, M.~H. et al.\ 
2004, in ASP Conf. Ser. 300, Radio Astronomy at the Fringe,\
ed. J.~A. Zensus, M.~H. Cohen, \& E. Ros, (San Francisco: ASP), 27

\bibitem[Denn, Mutel \& Marscher(2000)]{DMM00}
Denn, G.R., Mutel, R.L. \& Marscher, A.P. 2000, \apjs, 129, 61

\bibitem[de Vries, Barthel, \& O'Dea(1997)]{DBO97}
de Vries, W.~H., Barthel, P.~D., \& O'Dea, C.~P.\ 1997, \aap,
321, 105

\bibitem[Dondi \& Ghisellini(1995)]{DG95}
Dondi, L., \& Ghisellini, G.\
1995, \mnras, 273, 583


\bibitem[Giovannini et al.(2001)]{Gio01}
Giovannini, G., Cotton, W.~D., Feretti, L., Lara, L., \& Venturi, T.\
2001, \apj, 552, 508 

\bibitem[G\'omez et al.(2001)]{Gom01}
G\'omez, J.-L., Marscher, A.P., Alberdi, A., Jorstad, S.G., \& Agudo, I.\
2001, ApJ, 561, L161

\bibitem[Homan et al.(2001)]{Hom01}
Homan, D.C., Ojha, R., Wardle, J.F.C., Roberts, D.H.,
Aller, M.F., Aller, H.D. \& Hughes, P.A.\
2001, \apj, 549, 840

\bibitem[Homan et al.(2002)]{Hom02}
Homan, D.C., Ojha, R., Wardle, J.F.C., Roberts, D.H.,
Aller, M.F., Aller, H.D. \& Hughes, P.A.\
2002, \apj, 568, 99

\bibitem[Homan et al.(2003)]{Hom03}
Homan, D.~C., Lister, M.~L., Kellermann, K.~I., Cohen, M.~H.,
Ros, E., Zensus, J.~A., Kadler, M., \& Vermeulen, R.~C.\
2003, \apjl, 589, L9 

\bibitem[Homan et al.(2006)]{Hom06}
Homan, D.~C.; Kovalev, Y.~Y., Lister, M.~L., Ros, E., Kellermann, K. I.,
Cohen, M.~H., Vermeulen, R.~C., Zensus, J.~A., \& Kadler, M.\
2006, \apjl, 642, L115 

\bibitem[Jorstad et al.(2005)]{J05}
Jorstad, S.~G., et al.\ 
2005, \aj, 130, 1418
\citepalias{J05}

\bibitem[Jorstad et al.(2001)]{J01}
Jorstad, S. G., Marscher, A. P., Mattox, J. R., Wehrle, A. E.,
Bloom, S. D., \& Yurchenko, A. V.\ 2001, \apjs, 134, 181 

\bibitem[Kellermann et al.(1998)]{kik98}
Kellermann, K.~I., Vermeulen, R.~C., Zensus, J.~A., \& Cohen, M.~H.\
1998, \aj, 115, 1295
\citepalias{kik98}

\bibitem[Kellermann et al.(2004)]{Kellermann_etal04}
Kellermann, K.~I., Lister, M.~L., Homan, D.~C., Vermeulen, R.~C.,
Cohen, M.~H., Ros, E., Kadler, M., Zensus, J.~A., \& Kovalev, Y.~Y.\
2004, \apj, 609, 539
\citepalias{Kellermann_etal04}

\bibitem[Kovalev et al.(1999)]{KNK99}
Kovalev, Y.~Y., Nizhelsky, N.~A., Kovalev, Y.~A., Berlin, A.~B.,
Zhekanis, G.~V., Mingaliev, M.~G., \& Bogdantsov, A.~V.\
1999, \aaps, 139, 545

\bibitem[Kovalev et al.(2005)]{Ketal05}
Kovalev, Y.~Y., Kellermann, K.~I., Lister, M.~L., Homan, D. C.,
Vermeulen, R.~C., Cohen, M.~H., Ros, E., Kadler, M., Lobanov, A.~P.,
Zensus, J.~A., Kardashev, N.~S., Gurvits, L.~I., Aller, M.~F.,
\& Aller, H.~D.\
2005, \aj, 130, 2473
\citepalias{Ketal05}

\bibitem[Kraus et al.(2003)]{Kra03}
Kraus, A., et al.\
2003, \aap, 401, 161

\bibitem[L{\"a}hteenm{\"a}ki \& Valtaoja(1999)]{LV99} 
L{\"a}hteenm{\"a}ki, A., \& Valtaoja, E.\
1999, \apj, 521, 493 

\bibitem[Laing(1988)]{LAI88}
Laing, R.~A.\
1988, \nat, 331, 149

\bibitem[Lind \& Blandford(1985)]{LB85}
Lind, K.~R., \& Blandford, R.~D.\
1985, \apj, 295, 358 

\bibitem[Lister \& Homan(2005)]{LH05}
Lister, M.~L., \& Homan, D.~C.\
2005, \aj, 130, 1389 

\bibitem[Lister \& Marscher(1997)]{LM97}
Lister, M.~L., \& Marscher, A.~P.\
1997, \apj, 476, 572
\citepalias{LM97}

\bibitem[Lister et al.(2003)]{Lister_etal03}
Lister, M.~L., Kellermann, K.~I., Vermeulen, R.~C.,
Cohen, M.~H., Zensus, J.~A., \& Ros, E.\
2003, \apj, 584, 135 

\bibitem[Lobanov(1998)]{Lob98}
Lobanov, A.~P.\
1998, \aap, 330, 79 

\bibitem[Lovell et al.(2003)]{Lov03}
Lovell, J.~E.~J., Jauncey, D.~L., Bignall, H.~E., Kedziora-Chudczer, L.,
Macquart, J.-P., Rickett, B. J., \& Tzioumis, A.~K.\
2003, \aj, 126, 1699

\bibitem[O'Dea(1998)]{ODea98}
O'Dea, C.~P.\
1998, \pasp, 110, 493

\bibitem[Ogle et al.(1997)]{O97}
Ogle,P.M., Cohen, M.H., Miller, J.S., Tran, H.D \& Goodrich, R.W.\
1997, \apj, 482, L37


\bibitem[Piner et al.(2006)]{P06}
Piner, B.~G., Mahmud M., Fey, A.~L., \& Gospodinova, K.,\
2006, \aj, submitted

\bibitem[Taylor et al.(1996)]{TAY96}
Taylor, G.~B., Vermeulen, R.~C., Readhead, A.~C.~S., Pearson, T.~J.,
Henstock, D.~R., \& Wilkinson, P.~N.\
1996, \apjs, 107, 37

\bibitem[Ter{\"a}sranta et al.(2005)]{T05}
Ter{\"a}sranta, H., Wiren, S., Koivisto, P., Saarinen, V., \& Hovatta, T.\
2005, \aap, 440, 409

\bibitem[Urry \& Padovani(1995)]{UP95}
Urry, C.~M. \& Padovani, P.\
1995, PASP, 107, 803

\bibitem[Vermeulen \& Cohen(1994)]{VC94}
Vermeulen, R.~C., \& Cohen, M.~H.\
1994, \apj, 430, 467
\citepalias{VC94}

\bibitem[Vermeulen(1995)]{V95}
Vermeulen, R.~C.\
1995, PNAS, 92, 11385 

\bibitem[Vestergard \& Barthel(1993)]{VB93}
Vestergard, M., \& Barthel, P.D.\
1993, \aj, 105, 456

\bibitem[V{\'e}ron--Cetty \& V{\'e}ron(2003)]{VCV03}
V{\'e}ron-Cetty, M.-P., \& V{\' e}ron, P.\
2003, \aap, 412, 399 

\bibitem[Zensus et al.(2002)]{Z02}
Zensus, A., Ros, E., Kellermann, K.~I., Cohen, M.~H., 
Vermeulen, R.~C. \& Kadler, M. 2002, \aj, 124, 662 \citepalias{Z02}

\end{thebibliography}
\end{document}